\documentclass[applsci,article,accept,moreauthors,pdftex]{mdpi} 
\firstpage{1} 
\pubvolume{1}
\issuenum{1}
\articlenumber{0}
\pubyear{2021}
\copyrightyear{2020}
\datereceived{} 
\dateaccepted{} 
\datepublished{} 
\hreflink{https://doi.org/} 
\Title{Superfluid transition and specific heat of the 2D $x$-$y$ model: Monte Carlo simulation}
\TitleCitation{Title}
\Author{Phong Hai Nguyen  and Massimo  Boninsegni\orcidA{}*}
\AuthorNames{Phong H. Nguyen and Massimo Boninsegni}
\AuthorCitation{Nguyen, P. H.; Boninsegni, M.}
\address{\quad Department of Physics, University of Alberta, Edmonton, AB T6G 2E1, Canada}
\corres{Correspondence: m.boninsegni@ualberta.ca}
\abstract{We present results of large-scale Monte Carlo simulations of the 2D classical $x$-$y$ model on the square lattice. We obtain high accuracy results for the superfluid fraction and for the specific heat as a function of temperature, for systems of size $L\times L$ with $L$ up to $2^{12}$. Our estimate for the superfluid transition temperature is consistent with those furnished in all previous studies. The specific heat displays a well-defined peak, whose shape and position are independent of the size of the lattice for $L > 2^8$, within the statistical uncertainties of our calculations. The implications of these results on the interpretation of experiments on adsorbed thin films of $^4$He are discussed. }
\begin{document}
\section{Introduction}
The two-dimensional classical $x$-$y$ model is the simplest model to exhibit a Kosterlitz-Thouless (KT) transition \cite{Kosterlitz_1972,Kosterlitz_1973,Kosterlitz_1974,Nelson}. The KT universality class includes the superfluid phase transition in two dimensions (2D) which is a subject of ongoing experimental and theoretical investigation, chiefly in the context of thin films of $^4$He adsorbed on a wide variety of substrates \cite{bishop78,agnolet,csathy,toigo,taborek,kosterlitz}. Theoretical results obtained by studying the 
$x$-$y$ model, typically by computer simulations, are utilized both to ascertain whether a particular physical system experimentally investigated, believed to be in the same universality class, conforms with the KT paradigm, as well as to predict the behavior of systems yet unexplored \cite{dash78,moon,Schultka_1995,Schultka_1995b}. Decades of computer simulation studies of the 2D $x$-$y$ model, carried out on square lattices of size as large as $L=2^{16}$ \cite{Yukihiro_Komura},  have yielded very precise estimates of the superfluid transition temperature $T_c$  and of the critical exponents associated with the transition  \cite{Tobochnik,PhysRevB.23.359,Evertz,Gupta,PhysRevB.52.10221,Hasenbusch_2005,Yukihiro_Komura,sandvik,Youjin_Deng}.

Less extensively investigated is the behavior of the specific heat, which displays an anomaly at a temperature $\sim 17$\% above $T_c$, in numerical simulations of the $x$-$y$ model on square lattices of size $L=2^8$ \cite{Gupta}. The position of the peak appears to depend weakly on the size of the simulated lattice, but to our knowledge no systematic study has yet been carried out, aimed at establishing whether such an anomaly occurs in the thermodynamic limit, and its actual location. There have been also speculations that the width of the peak may narrow in the thermodynamic limit, and the peak itself may evolve into a cusp \cite{PhysRevB.23.359}. There is presently no consensus regarding the physical interpretation of such an anomaly, which does not appear to signal the occurrence of any phase transition. Interestingly, experiments on $^4$He monolayers \cite{PhysRevLett.71.3673}, as well as  computer simulations \cite{PhysRevB.102.235436} (including of 2D $^4$He \cite{PhysRevB.39.2084}) have also yielded evidence of a peak in the specific heat at temperature {\em above} the superfluid transition temperature.

To our knowledge no further studies have been carried out of the specific heat, beyond that of Ref. \cite{Gupta}, aimed at establishing any possible shift in temperature of the peak, as the lattice size is increased, as well as the general shape of the curve. One reason for this state of affairs is that the calculation of the specific heat in direct numerical (Monte Carlo) simulations is affected by relative large statistical uncertainties, due to the inherent ``noisiness" of the presently known  specific heat estimators. However, the (almost) three decades elapsed since the publication of Ref. \cite{Gupta} have witnessed both an impressive increase in the available computing power, as well as the development of more efficient and sophisticated simulation methods. In light of that, it seems worthwhile to revisit this issue, which is of potential experimental relevance, as the $x$-$y$ model is sometimes invoked in the interpretation of measurements of the specific heat of thin $^4$He films, as well as utilized predictively, in the same context \cite{Schultka_1995,Schultka_1995b}. 

In this article, we report results of large scale computer simulations of the 2D $x$-$y$ model on the square lattice, performed using the Worm Algorithm in the lattice flow representation \cite{Prokofev}. We carried out simulations on square lattices of size up to $L=2^{12}$. 
Our main aim is to study the specific heat, and provide robust, reliable information about its behavior in the thermodynamic limit; in order to validate our study, we also computed the superfluid transition temperature and spin correlations, comparing them to the most recent theoretical estimates. We obtain a value of  $T_c$ consistent with that of Ref. \cite {Yukihiro_Komura}, which is presently the most accurate published result. We present strong numerical evidence confirming the presence of the specific heat anomaly in the thermodynamic limit, its shape remaining essentially unchanged with respect to that on a lattice of size $L=2^8$. We estimate the position of the peak of the specific heat in the thermodynamic limit to be at temperature 1.043(4) (in units of the coupling constant).

This paper is organized as follows. In Sec. \ref{Sec.2}, we briefly describe our computational methodology, in Sec. \ref{Sec.3}, we analyze the MC data and present the results. We outline our conclusions in Sec. \ref{concl}.
 
\section{Model and Methodology}
\label{Sec.2}
The Hamiltonian of the classical $x$-$y$ model is given by
\begin{equation}
    H = -J\sum_{\langle ij\rangle}{\bf s}_i\cdot{\bf s}_j
\end{equation}
where the sum runs over all pair of nearest-neighboring sites, and  $\textbf{s}_i\equiv s(\cos{\theta_{i}},\ \sin{\theta_{i}})$ is a classical spin variable associated with site $i$. We assume a square lattice of $N=L\times L$ sites, with periodic boundary conditions. Henceforth, we shall take our energy (and temperature) unit to be $Js^2$.
We investigate the low-temperature physics of this model by carrying out classical Monte Carlo simulations, based on the methodology mentioned above, which is extensively described in the original reference \cite{Prokofev} and will therefore not be reviewed here. Details of our calculations are standard.

As mentioned above, an important part of this work consists of studying the superfluid transition, in order to compare our results with those of existing studies and gauge therefore the accuracy of our methodology. We determine the superfluid transition temperature $T_c$ in two different ways. The first consists of computing the superfluid fraction  $\rho_s(L,T)$ on a lattice of size $L$, as a function of temperature, using the well-known winding number estimator \cite{Pollock}. We then determine a size-dependent transition temperature $T_c(L)$ based on the universal jump condition \cite{Nelson_Kosterlitz}
\begin{align}
\rho_{s}(L,T_c)=f_r\ \frac{2T_{c}}{\pi}
\label{Eq2.3}
\end{align}
where $f_r=1-16\pi e^{-4\pi}$ \cite{Prokofev_2000}. Eq. \ref{Eq2.3} can be used to obtain an estimate of the transition temperature ($T_c(L)$) on a lattice of finite size.  In order to extrapolate the value of $T_c (L)$ to the thermodynamic ($L\to\infty$) limit (referred to as $T_c$), we fit the results for $T_c(L)$, obtained for different system sizes to the expression \cite{Yusuke_Tomita}:

\begin{align}
T_c(L)=T_c+\frac{a}{(\text{ln}bL)^2}
\label{Eq2.4}    
\end{align}
where $a$, $b$ are constant. It should be noted that other expressions have been proposed, aimed at extracting $T_c$ \cite{Youjin_Deng}; we come back to this point when discussing our results.

The superfluid transition temperature can also be inferred from the behavior of the spin correlation function \cite{Janke}, specifically from the divergence of the correlation length $\xi$ as $T\to T_c$, namely 
\cite{Kosterlitz_1972,Kosterlitz_1973,Kosterlitz_1974}:
\begin{align}
\xi(T)\sim A\ e^{\frac{c}{\sqrt{t}}}
\label{Eq2.5}    
\end{align}
where $A$, $c$ are constant ($c\approx1.5$ \cite{Janke}), and $t=\frac{(T-T_{c})}{T_{c}}$ is the reduced temperature. Above $T_c$, the correlation length $\xi(T)$ can be obtained from the computed correlation function by means of a simple fitting procedure, illustrated in Ref. \cite{Tobochnik}. Using the best fit to Eq. (\ref{Eq2.5}), an estimate of $T_{c}$ is obtained; the accuracy of the estimated $T_c$ increases with the size of the system studied.
The estimates of $T_c$ obtained in the two ways illustrated above are consistent within their statistical uncertainties; however, we find that the first procedure, based on the universal jump of the superfluid fraction, affords  a considerably more accurate determination of $T_c$. 

Moreover, we calculate the specific heat (i.e., the heat capacity per site) through the direct estimator of the heat capacity \cite{GouldandTobochnik}, based on the mean-squared fluctuations of the total energy $E$:
\begin{align}
C=\frac{1}{L^2}\beta^2(\langle E^2\rangle-\langle E\rangle^2),
\label{Eq2.8}    
\end{align}
{\color {black} where $\beta=1/T$ is the inverse temperature}.
This estimator is numerically ``noisy'', and for this reason the numerical differentiation of the computed energy values with respect to the temperature has often been preferred \cite{Tobochnik}. In our case, however, we found it possible to obtain reasonably accurate estimates of the specific heat using Eq. (\ref{Eq2.8}), thanks to the available computing facilities and the methodology adopted.

\section{Results}
\label{Sec.3}
\begin{figure}[h]
\centering
\includegraphics[scale=0.4]{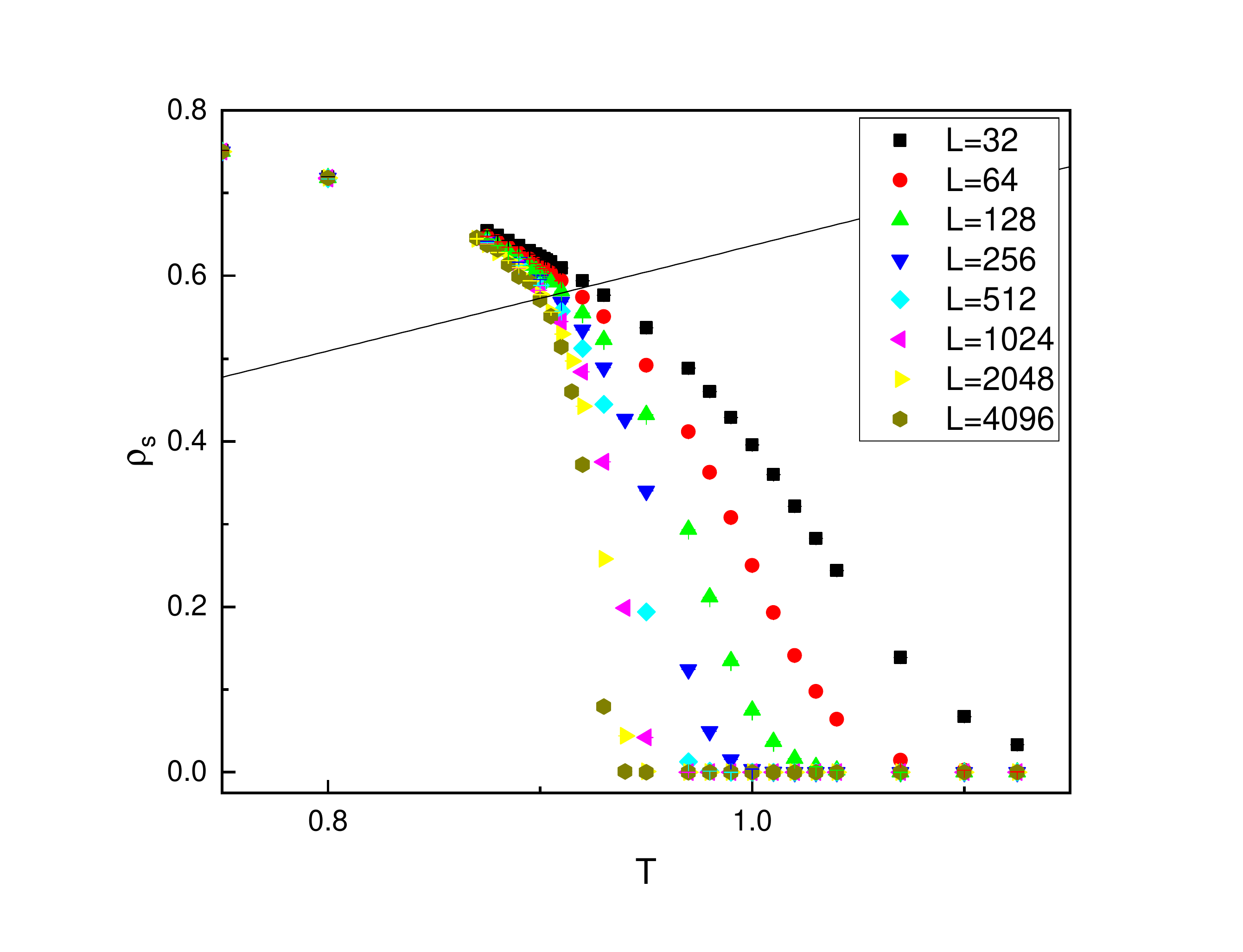}
\caption{The superfluid fraction $\rho_s$ versus  temperature, for the different lattice sizes considered. Statistical errors are smaller than symbol sizes. The straight line corresponds to the universal jump condition (right hand side of Eq. \ref{Eq2.3})}
\label{Fig.1}
\end{figure}

\begin{figure}[h]
\centering
\includegraphics[scale=0.45]{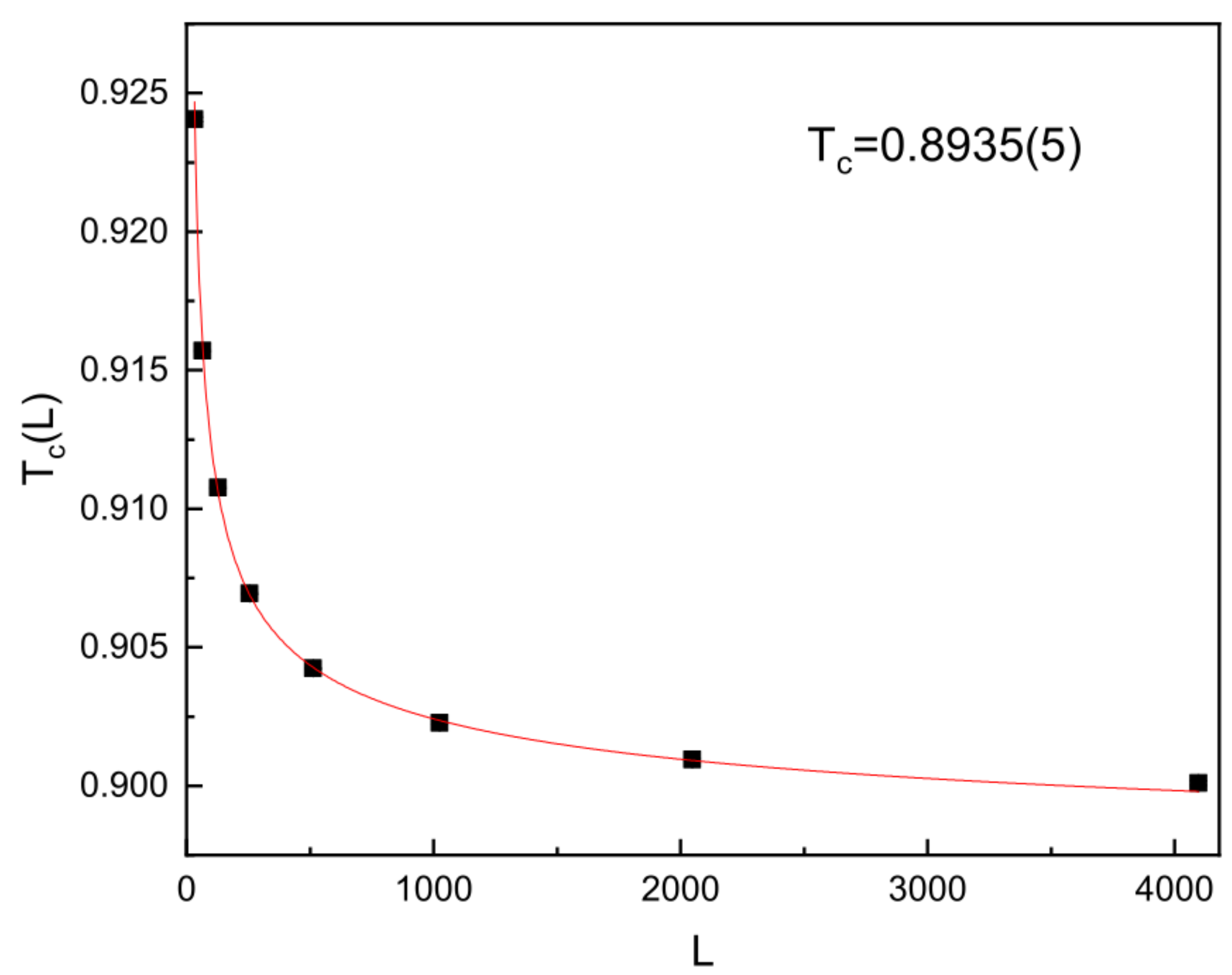}
\caption{The critical temperature $T_c(L)$ versus the system size $L$. Solid line is a fit to the data using Eq. \ref{Eq2.4}.}
\label{Fig.2}
\end{figure}
We begin  the presentation of our results by illustrating our estimates for the superfluid fraction as a function of temperature for the various lattice sizes considered, and by discussing the determination of the transition temperature, which we compare to those provided in other works.
Fig. \ref{Fig.1} shows the computed value of $\rho_s(L,T)$; the critical temperature $T_c(L)$ for a given system size is determined  by the universal jump condition, namely the intersection of the $\rho_s(L,T)$ curve with the straight line given by the right hand side of Eq. \ref{Eq2.3}.  The intersection point is estimated by drawing a straight line between the two adjacent values of $\rho_s(L,T)$ within which the intersection can be established to take place, within the precision of our calculation. 

As expected, both $T_c(L)$ and $\rho_s(L,T_c(L))$ display a slow decrease as a function of $L$. In order to extrapolate the value of $T_c$  in the thermodynamic ($L\to\infty$) limit, we fit the computed $T_c(L)$ to Eq. (\ref{Eq2.4}), as proposed in Ref. \cite{Yukihiro_Komura}. This procedure is illustrated in Fig. \ref{Fig.2}. Our estimate is $T_c=0.8935(5)$, which is consistent with that of Ref. \cite{Yukihiro_Komura}, namely 0.89289(5), even though their quoted uncertainty is a factor ten smaller than ours, a fact to be ascribed to the  significantly (sixteen times) greater system sizes adopted therein. Our estimate of $T_c$ is also in perfect agreement with the more recent result of Ref. \cite{Youjin_Deng}, in which the same computational methodology utilized in this work was adopted, on the same system sizes. Their results are of precision comparable to ours; they make use of a different, more elaborate fitting form for $T_c(L)$, but their resulting estimate for $T_c$ is entirely consistent with ours, and has the same uncertainty.

\begin{figure}[h]
\centering
\includegraphics[scale=0.45]{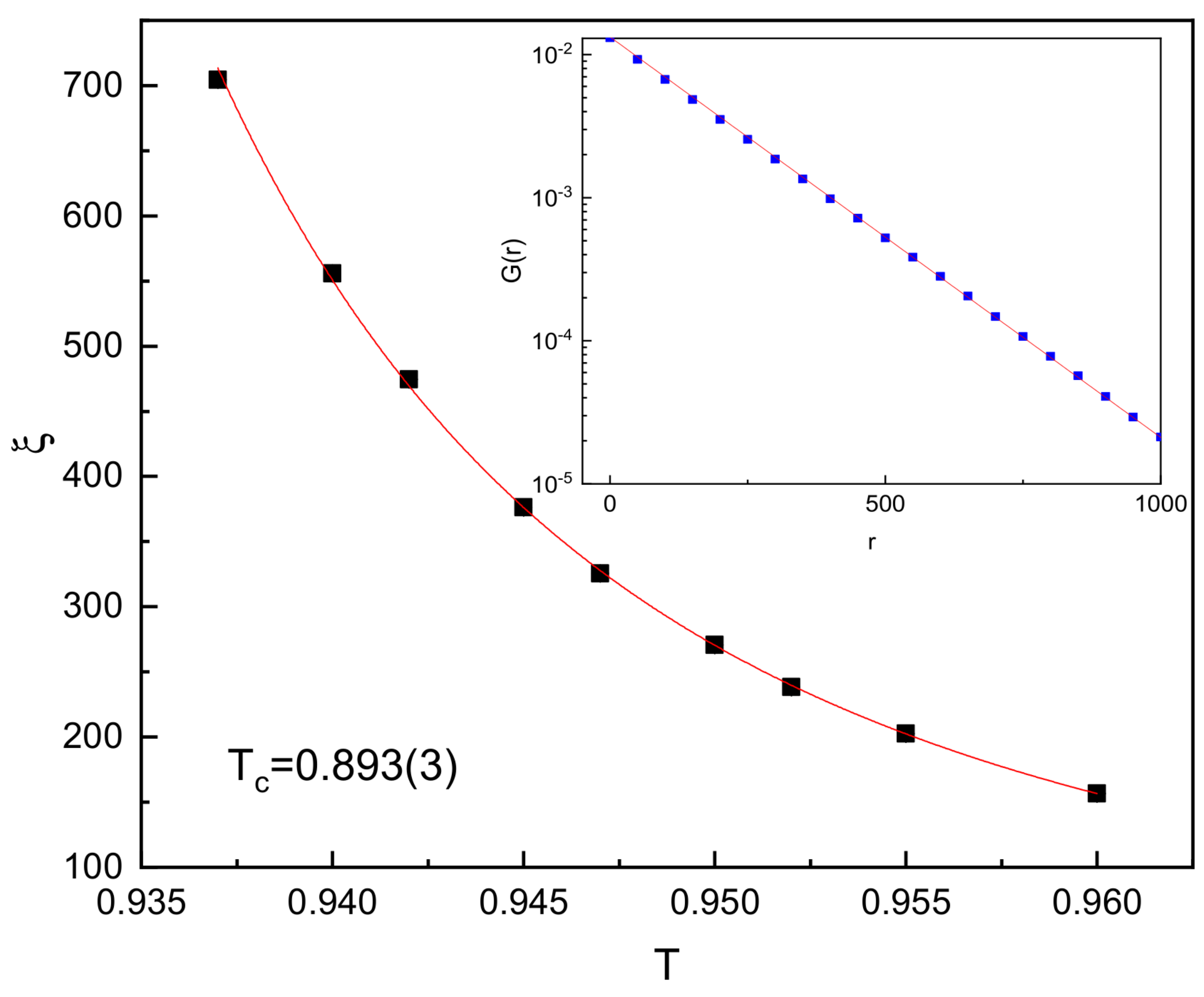}
\caption{The correlation length $\xi$ as a function of the temperature, for a system of size $L=4096$. The solid line is a fit to the data using expression (\ref{Eq2.5}). Inset shows the computed spin correlation function $G(r)$ for a temperature $T=0.96$. }
\label{Fig.3}
\end{figure}

As mentioned in Sec. \ref{Sec.2}, as a further check of our results we estimated the critical temperature $T_c$ independently through the spin correlation length. In this case, the computed spin correlation function for a given system size yields an estimate of $T_c$, obtained first by extracting the correlation length $\xi(T)$ as a function of temperature, and then fitting the results to Eq. \ref{Eq2.5}. An example of this procedure is shown in Fig. \ref{Fig.3}, for the largest system size considered here, which is the one that yields the estimate of $T_c$ with the smallest uncertainty. Such an estimate, namely 0.893(3), is consistent with that obtained from the superfluid fraction, but it is considerably less accurate.
\begin{figure}[h]
\centering
\includegraphics[scale=0.5]{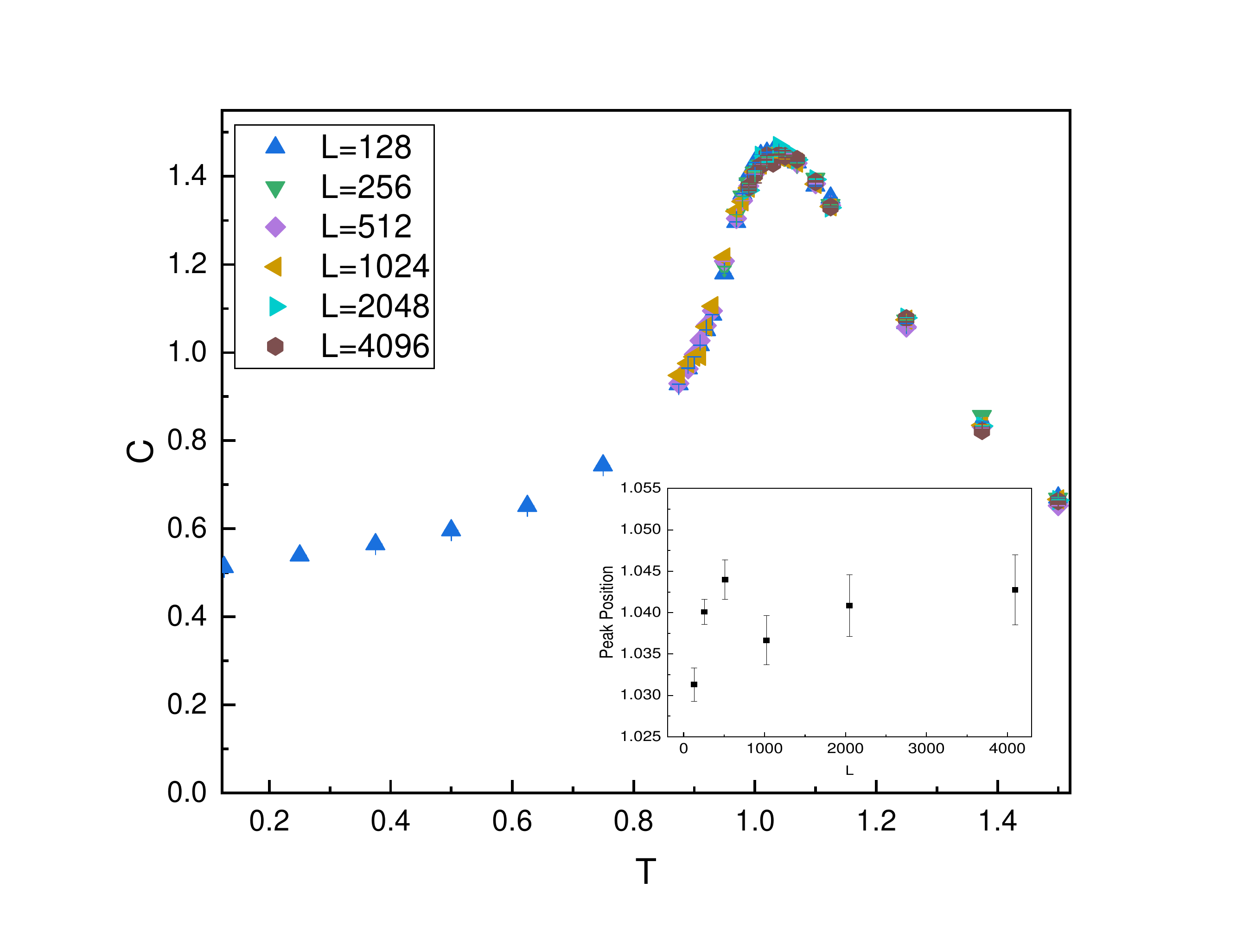}
\caption{The specific heat $C$ versus the temperature $T$ for different lattice sizes. The inset shows the position of the peak as a function of lattice size.}
\label{Fig.4}
\end{figure}
Our result for $T_c$ gives us sufficient confidence on the reliability of our data and simulation. Therefore, we now discuss the most important part of this work, namely the behavior of the specific heat $C(T)$. It is worth restating that previous numerical studies of the 2D $x$-$y$ model \cite{Tobochnik,Gupta} have yielded results for this quantity only for square lattices of size up to $L=256$. Such studies yielded evidence of a peak in the specific heat at a temperature above $T_c$; the position of this peak depends fairly strongly on system size for $L\leq128$. On the other hand, the shape of $C(T)$ appears to change little going from $L=128$ to $L=256$, suggesting that the anomaly may indeed be a genuine physical feature of the model and not an artifact of numerical simulations carried out on finite systems of small size.

Fig. \ref{Fig.4} shows our results for the specific heat for the various system sizes, showing that the curve indeed appears to stabilize for $L > 256$. The inset of Fig. \ref{Fig.4} shows the position of the peak, which, within the statistical uncertainties of our calculation, is independent on system size. Our best estimate of the peak position is $T_P=1.043(4)=1.167(1)\ T_c$.  The height of the peak is approximately 1.45. Altogether, therefore, our simulations aimed at characterizing quantitatively  the specific heat anomaly, carried out on lattices of significantly greater size than those of all previous studies (of the specific heat), revise the position of the peak to a slightly higher temperature, and reduce its height by a few percent. However, the presence of the anomaly, its overall shape, the fact that it remains broad (i.e., it does not approach a cusp in the thermodynamic limit) and that it occurs at a different temperature than the superfluid transition, can in our view be regarded at this point as well-established. It is worth reminding that the presence of such an anomaly has been theoretically predicted using different approaches, furnishing results in quantitative agreement with those of Monte Carlo simulations \cite{Jakubczyk}. 

\section{Conclusions}\label{concl}
Summarizing, we have carried out extensive Monte Carlo simulations of the 2D $x$-$y$ model, making use of the Worm Algorithm. The twofold purpose of our study was that of assessing the effectiveness of the methodology, which to our knowledge has not yet been applied this particular model (we became aware of Ref. \cite{Youjin_Deng} while this work was in progress), as well as consolidating existing theoretical results for the specific heat. We have simulated the model on lattices of linear size up to $L=4096$, obtaining for the superfluid transition temperature results of accuracy comparable to that yielded by the most recent numerical simulations, making use of standard computational resources.
For the specific heat, the largest system size for which we report results is sixteen times greater than that for which Monte Carlo estimates have been published. Our results confirm the existence of a specific heat anomaly, namely a peak, occurring at a temperature $\sim 17\%$ higher than that at which the superfluid transition temperature takes place. It is interesting to compare this to 2D $^4$He, for which the peak in the specific heat observed in computer simulations \cite{PhysRevB.39.2084} is located at $T\sim 1.6\ T_c$.

It has been suggested \cite{Tobochnik,Ota,ISI:A1995RX18900022} that the temperature dependence of the specific heat correlates with that of the vortex density above the critical temperature. In this case, one could expect a similar specific heat anomaly, which is not indicative of a phase transition, to occur in physical systems such as atomically thin $^4$He films, which approach the 2D limit and display superfluid transitions that conform with the KT paradigm. Indeed, this may help in the interpretation of specific heat data for $^4$He films adsorbed on graphite, where similar features (peaks) are often interpreted as signalling phase transitions (e.g., melting of commensurate solid phases, see for instance Ref. \cite{PhysRevB.102.235436}).

\authorcontributions{The authors contributed equally to this work.}

\funding{This research was funded by the Natural Science and Engineering Research Council of Canada. Computing support from ComputeCanada is gratefully acknowledged.}

\dataavailability{The computer codes utilized to obtain the results can be obtained by contacting the authors.} 

\acknowledgments{The authors wish to thank Prof. Y. Deng for sharing the results of his independent investigation.}

\conflictsofinterest{The authors declare no conflict of interest.} 
\end{paracol}
\reftitle{References}
\externalbibliography{yes}
\bibliography{sample}
\end{document}